\documentstyle[11pt]{article}
\begin{document}
\baselineskip=0.43cm
\twocolumn[\hsize\textwidth\columnwidth\hsize\csname
@twocolumnfalse\endcsname

\title{{\large{\bf Do we need photons in parametric down
conversion?}}}
\author{
Trevor~W.~Marshall\\
{\small{\it Department of Mathematics, University of Manchester,
Manchester M13 9PL, U. K.}}}
\date{\today}
\maketitle

\begin{abstract}
The phenomenon of parametric
down conversion from the vacuum may be understood as a process
in classical electrodynamics, in which a nonlinear
crystal couples the modes of the pumping
field with those of the zeropoint, or ``vacuum", field.
This is an entirely local theory of the
phenomenon, in contrast with the
presently accepted nonlocal theory.
The new
theory predicts a hitherto unsuspected phenomenon --- parametric
up conversion
from the vacuum.\\
\vspace{0.5cm}

\end{abstract}

\vskip2pc]

The widely accepted description of parametric\\
down conversion (PDC) by a pumped nonlinear
crystal (NLC) is based on a hamiltonian (for simplicity
we confine attention to the scalar version)
\begin{equation}
H=\sum \hbar\omega_{\bf k}a_{\bf k}^{\dagger}a_{\bf k}
+\sum g_{{\bf kk'}}a^\dagger_{\bf k}a^\dagger_{\bf k'}
e^{i\omega_0t}\;.
\end{equation}
The interaction part of $H$ is interpreted to mean that
(I prefer to say ``was constructed so that")
an incoming laser photon of frequency $\omega_0$
down converts into a correlated pair of photons
with frequencies $\omega$ and $\omega_0-\omega$. Such
a process is depicted in Fig.1.
\begin{figure}[htb]
\unitlength=0.55mm
\linethickness{0.4pt}
\begin{picture}(159.33,44.67)
\put(70.00,44.67){\line(0,-1){40.00}}
\put(90.00,4.67){\line(0,1){40.00}}
\thicklines
\put(0.00,24.67){\line(1,0){70.00}}
\thinlines
\put(90.00,24.67){\line(5,1){69.33}}
\put(159.33,10.67){\line(-5,1){69.33}}
\put(80.00,24.67){\makebox(0,0)[cc]{{NLC}}}
\put(80.00,31.67){\makebox(0,0)[cc]{{}}}
\put(80.00,17.67){\makebox(0,0)[cc]{{}}}
\put(12.67,24.67){\vector(1,0){7.67}}
\put(119.33,18.67){\vector(4,-1){1.00}}
\put(119.67,30.67){\vector(4,1){1.00}}
\end{picture}

\caption{\small{PDC --- photon-theoretic version. A laser photon
down converts into a conjugate pair of PDC photons
with conservation of energy.}}
\end{figure}
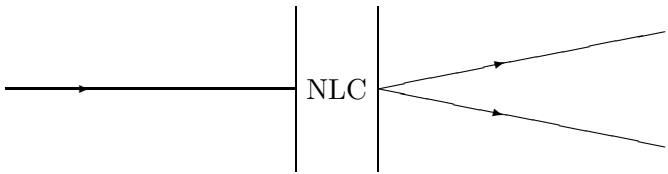
Naturally, since we know that $E=\hbar\omega$,
that means energy is conserved in the PDC process.
The correlation in the properties of the two outgoing
photons, according to this description, can only be
described as bizarre\cite{zwm,kwiat}.
Indeed, attempts,
by supporters of the theory, to describe effects such
as ``entanglement" to a nonspecialist public, have used the
words ``mind boggling"\cite{ghz} and, in
another context, ``absurd"\cite{bz},
but these authors' implication is that the theory
must nevertheless be correct.

The mind boggling feature of the standard
theory of PDC, and also of atomic transitions,
is its nonlocality. This feature
is already clear in Fig.1, which implies that
photons, or plane waves filling the whole of space,
are created instantaneously by the action of $H$.
But we do not have to believe Fig.1 represents
what really happens. The alternative to photons
is almost as old as photons themselves, namely
the zeropoint electromagnetic field of Max
Planck\cite{milonni,mex}.

I have reviewed
elsewhere\cite{vig} how
nonlocal photon descriptions may be more or
less systematically replaced by local descriptions
which incorporate Planck's field. This systematic
replacement includes, indeed begins with, the
celebrated experiments on Bell-\\
inequality tests
in atomic cascades\cite{marshsant}. Most specialists,
possibly quite rightly, are not impressed, because,
although the treatment of the {\it field} is
unambiguously maxwellian, we have had to improvise,
in a rather crude manner, the description of
the atom-field interaction. Nobody has yet
succeeded in doing, for the Dirac field, what
Planck did for the Maxwell field.

Now when it comes to PDC we do not need to
know any details about the atom-field interaction;
only the relation between the current and the
field inside the crystal is relevant at optical and
near ultraviolet frequencies. So a purely
maxwellian theory of the type I have
outlined\cite{vig,magic}, and which I call the Field Theory,
can engage with the theory generated by eq.(1) and Fig.1,
which I call the Photon Theory,
on equal terms.
But the Field Theory is manifestly
local and causal; the incoming fields generate
a current in the crystal, and the outgoing fields
are the retarded fields generated by those currents.

The form of maxwellian theory I have considered
starts from the current-field relation
\[
J({\bf r},t)= \int_{-\infty}^\infty
f(\omega)d\omega\int_{-\infty}^\infty
E({\bf r},t')e^{i\omega t-i\omega t'}dt'
\]
\begin{equation}
+g\cos[\omega_0t-\omega_0\mu(\omega_0)z]E({\bf r},t)\;,
\label{maxcoup}
\end{equation}
where  $f(\omega)$ is analytic in the lower half plane,
so that the integration on $t'$ may be taken from
minus infinity to $t$ only.
The function $f(\omega)$ is related to the refractive index
$\mu(\omega)$ in the standard way, and it is this function
which determines the directions of the
down converted waves.
Eq.(2), like eq.(1), is a linearized form of the
interaction, that is it neglects effects due to depletion of
the laser intensity.

We have shown, in a series of
articles\cite{pdc1,pdc2,pdc3,pdc4}, that the
Photon Theory, when cast in the Wigner representation,
gives a coupling between the field modes. The
PDC predictions of that theory may then all be
obtained by the application of {\it second-order}
perturbation theory based on $H$, as given by eq.(1).
The Hilbert-space representation is
generally preferred, because in that case only
{\it first-order} perturbation theory is required,
but I think this algorithmic simplification
obscures what is really happening inside the
crystal. I say this because the field-coupling
obtained from eq.(\ref{maxcoup}) is {\it very nearly}
the same as we obtained in the above series of articles,
and a calculation of one important set of data,
namely the fringe visibility in the Zou-Wang-Mandel
experiment\cite{zwm}, using second-order perturbation
theory with this coupling, gives {\it very nearly} the
same results as the Photon Theory.

The point about first- and second-order contributions
is made clear by reference to Fig.2, which may be
considered a picture of classical PDC\cite{saleh}.
In order to find the intensities of the outgoing
waves to second order, we need the signal
amplitude to first order, {\it and the idler amplitude
to second order}.
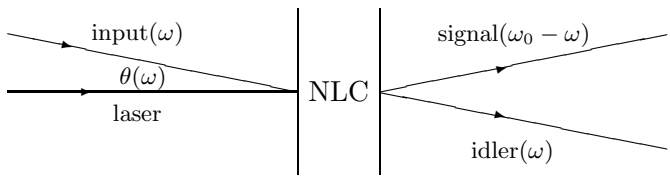
\begin{figure}[htb]
\unitlength=0.55mm
\linethickness{0.4pt}
\begin{picture}(159.33,44.67)
\put(70.00,44.67){\line(0,-1){40.00}}
\put(90.00,4.67){\line(0,1){40.00}}
\thicklines
\put(0.00,24.67){\line(1,0){70.00}}
\thinlines
\put(70.00,24.67){\line(-5,1){70.00}}
\put(90.00,24.67){\line(5,1){69.33}}
\put(159.33,10.67){\line(-5,1){69.33}}
\put(31.33,19.67){\makebox(0,0)[cc]{{\footnotesize laser}}}
\put(31.33,38.67){\makebox(0,0)[cc]{{\footnotesize input$(\omega)$}}}
\put(122.33,38.34){\makebox(0,0)[cc]{{\footnotesize signal$(\omega_0-\omega)$}}}
\put(122.00,10.00){\makebox(0,0)[cc]{{\footnotesize idler$(\omega)$}}}
\put(80.00,24.67){\makebox(0,0)[cc]{NLC}}
\put(80.00,31.67){\makebox(0,0)[cc]{{}}}
\put(80.00,17.67){\makebox(0,0)[cc]{{}}}
\put(12.67,24.67){\vector(1,0){7.67}}
\put(14.67,35.67){\vector(4,-1){1.00}}
\put(119.33,18.67){\vector(4,-1){1.00}}
\put(119.67,30.34){\vector(4,1){1.00}}
\put(32.67,28.00){\makebox(0,0)[cc]{{\footnotesize $\theta(\omega)$}}}
\end{picture}

\caption{\small{Classical PDC. When a wave of
frequency $\omega$ is incident, at a certain angle
$\theta(\omega)$, on a nonlinear crystal
pumped at frequency $\omega_0$, a signal
of frequency $\omega_0-\omega$ is emitted
in a certain conjugate direction. The modified
input wave is called the idler.}}
\end{figure}

There is, of course, a symmetry about Fig.1 which is
lacking
in Fig.2. This is because we have, in Fig.2, no input
corresponding to the signal mode. But the
point about our Planck-based theory is that {\it all}
modes of the field are actually present, with an intensity
corresponding to half a photon. If one input is
classical, for example a second laser, it is
appropriate to leave out its conjugate mode as we have done.
But if we wish to consider PDC {\it from the vacuum},
which is what Fig.1 purports to describe, we have
to take both relevant inputs into account, as in Fig.3.
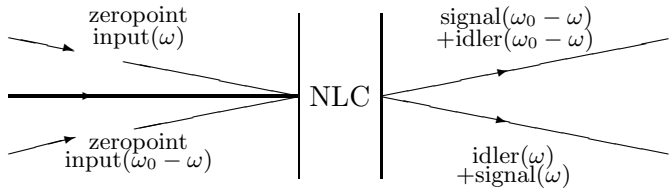
\begin{figure}[htb]
\unitlength=0.55mm
\linethickness{0.4pt}
\begin{picture}(159.33,44.67)
\put(70.00,44.67){\line(0,-1){40.00}}
\put(90.00,4.67){\line(0,1){40.00}}
\thicklines
\put(0.00,24.67){\line(1,0){70.00}}
\thinlines
\put(31.67,28.34){\makebox(0,0)[cc]{{}}}
\put(90.00,24.67){\line(5,1){69.33}}
\put(159.33,10.67){\line(-5,1){69.33}}
\put(11.33,19.67){\makebox(0,0)[cc]{}}
\put(31.33,38.67){\makebox(0,0)[cc]{{\footnotesize input($\omega$)}}}
\put(122.33,38.34){\makebox(0,0)[cc]{{\footnotesize +idler($\omega_0-\omega$)}}}
\put(122.00,10.00){\makebox(0,0)[cc]{{\footnotesize idler($\omega$)}}}
\put(80.00,24.67){\makebox(0,0)[cc]{NLC}}
\put(80.00,31.67){\makebox(0,0)[cc]{{}}}
\put(80.00,17.67){\makebox(0,0)[cc]{{}}}
\put(12.67,24.67){\vector(1,0){7.67}}
\put(14.67,36.00){\vector(4,-1){1.00}}
\put(119.33,18.67){\vector(4,-1){1.00}}
\put(119.67,30.67){\vector(4,1){1.00}}
\put(70.00,24.67){\line(-5,1){42.33}}
\put(15.67,35.67){\line(-5,1){15.67}}
\put(0.00,38.13){\line(0,0){0.00}}
\put(31.33,44.00){\makebox(0,0)[cc]{{\footnotesize zeropoint}}}
\put(122.67,43.67){\makebox(0,0)[cc]{{\footnotesize signal($\omega_0-\omega$)}}}
\put(122.00,5.33){\makebox(0,0)[cc]{{\footnotesize +signal($\omega$)}}}
\put(0.00,10.67){\vector(4,1){16.00}}
\put(70.00,24.67){\line(-5,-1){39.67}}
\put(31.33,13.33){\makebox(0,0)[cc]{{\footnotesize zeropoint}}}
\put(31.33,9.00){\makebox(0,0)[cc]{{\footnotesize
input($\omega_0-\omega$)}}}
\put(33.33,21.67){\makebox(0,0)[cc]{{}}}
\end{picture}

\caption{\small{PDC from the vacuum --- field-theoretic version. Both of
the outgoing signals are above zeropoint intensity, and
hence give photomultiplier counts.}}
\end{figure}
The zeropoint inputs, denoted by interrupted lines,
do not activate photodetectors, because the
threshold of these devices is set precisely at the
level of the zeropoint intensity, as discussed in
Ref.\cite{pdc4}. However,
the two idlers have intensities above
that of their corresponding inputs.
Also there is no
coherence between a signal and an idler of the same
frequency, so  their intensities
are additive in both channels. Hence there are photoelectron
counts in both of the outgoing channels of Fig.3.

The small differences between the predictions of eqs.(1) and (2),
even for all PDC processes considered hitherto,
let alone just the one or two I have already tested, does not appear
to be a convincing reason to change our allegiance from (1) to
(2); the day-to-day practice of science is such that we
do not need a reason for staying with (1), even though
the latter is now admitted\cite{ghz,bz} to be absurd.
Students of history will tell us we need a crucial experiment\ldots\\

Right then. Here it is.

An incident wave
of frequency $\omega$, as well as being
down converted by the pump to give
a PDC signal of frequency $\omega_0-\omega$,
may also be {\it up converted to give a
PUC signal} of frequency $\omega_0+\omega$.
We depict this phenomenon, which is well
known\cite{saleh,yariv} in classical nonlinear optics,
in Fig.4.
Note that the angle of incidence, $\theta_u(\omega)$,
at which PUC occurs is quite different from the PDC
angle, which in Fig.2 was denoted simply $\theta(\omega)$,
but which we should now call $\theta_d(\omega)$.
\begin{figure}[h]
\unitlength=0.55mm
\linethickness{0.4pt}
\begin{picture}(159.33,86.80)
\put(70.00,85.00){\line(0,-1){80.00}}
\put(90.00,5.00){\line(0,1){80.00}}
\thicklines
\put(0.00,45.00){\line(1,0){70.00}}
\thinlines
\put(80.00,45.00){\makebox(0,0)[cc]{{NLC}}}
\put(80.00,52.00){\makebox(0,0)[cc]{{}}}
\put(80.00,38.00){\makebox(0,0)[cc]{{}}}
\put(12.67,45.00){\vector(1,0){7.67}}
\put(70.00,45.00){\line(-5,3){69.67}}
\put(90.00,45.00){\line(5,-3){69.33}}
\put(159.33,3.40){\line(0,0){0.00}}
\put(90.00,45.00){\line(6,-1){69.33}}
\put(49.00,51.00){\makebox(0,0)[cc]{{\footnotesize $\theta_u(\omega)$}}}
\put(32.33,38.67){\makebox(0,0)[cc]{{\footnotesize laser}}}
\put(33.67,73.67){\makebox(0,0)[cc]{{\footnotesize input($\omega$)}}}
\put(131.67,43.33){\makebox(0,0)[cc]{{\footnotesize signal($\omega_0+\omega$)}}}
\put(115.67,21.00){\makebox(0,0)[cc]{{\footnotesize idler($\omega$)}}}
\put(120.00,27.00){\vector(3,-2){1.00}}
\put(121.67,39.67){\vector(4,-1){1.00}}
\put(27.67,70.33){\vector(3,-2){1.00}}
\end{picture}

\caption{\small{PUC. In contrast with PDC
the output signal has its
transverse component in the same direction as that of
the idler.}}
\end{figure}
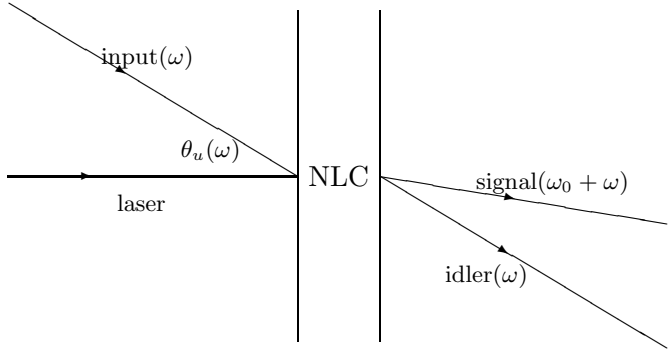

Now, following the same argument which led us from
Fig.2 to Fig.3, we predict the phenomenon of
PUC from the vacuum,
which we depict in Fig.5.
When we come to calculate
the intensity of the PUC rainbow\cite{puc1},
there is an important
\begin{figure}[htb]
\unitlength=0.55mm
\linethickness{0.4pt}
\begin{picture}(159.33,85.00)
\put(70.00,85.00){\line(0,-1){80.00}}
\put(90.00,5.00){\line(0,1){80.00}}
\thicklines
\put(0.00,45.00){\line(1,0){70.00}}
\thinlines
\put(80.00,45.00){\makebox(0,0)[cc]{{NLC}}}
\put(80.00,52.00){\makebox(0,0)[cc]{{}}}
\put(80.00,38.00){\makebox(0,0)[cc]{{}}}
\put(12.67,45.00){\vector(1,0){7.67}}
\put(90.00,45.00){\line(5,-3){69.33}}
\put(159.33,3.40){\line(0,0){0.00}}
\put(32.33,38.67){\makebox(0,0)[cc]{{\footnotesize laser}}}
\put(33.67,73.67){\makebox(0,0)[cc]{{\footnotesize input($\omega$)}}}
\put(131.67,43.33){\makebox(0,0)[cc]{{\footnotesize +idler($\omega_0+\omega$)}}}
\put(115.67,21.00){\makebox(0,0)[cc]{{\footnotesize idler($\omega$)}}}
\put(120.00,27.00){\vector(3,-2){1.00}}
\put(121.67,39.67){\vector(4,-1){1.00}}
\put(27.67,70.00){\vector(3,-2){1.00}}
\put(33.33,79.67){\makebox(0,0)[cc]{{\footnotesize zeropoint}}}
\put(70.00,45.00){\line(-6,1){30.67}}
\put(18.00,54.00){\line(-6,1){18.00}}
\put(13.00,55.00){\vector(4,-1){1.00}}
\put(25.00,56.33){\makebox(0,0)[cc]{{\footnotesize input($\omega_0+\omega$)}}}
\put(25.00,61.33){\makebox(0,0)[cc]{{\footnotesize zeropoint}}}
\put(131.33,50.00){\makebox(0,0)[cc]{{\footnotesize signal($\omega_0+\omega$)}}}
\put(115.67,14.67){\makebox(0,0)[cc]{{\footnotesize +signal($\omega$)}}}
\put(70.00,45.00){\line(-5,3){27.00}}
\put(28.67,69.33){\line(-5,3){22.33}}
\put(90.00,45.00){\line(6,-1){32.33}}
\put(137.33,36.33){\line(6,-1){21.33}}
\end{picture}

\caption{\small{PUC from the vacuum.  Only one
of the outgoing signals is above the zeropoint intensity. The
other one, depicted by an interrupted line, is below
zeropoint intensity.}}
\end{figure}
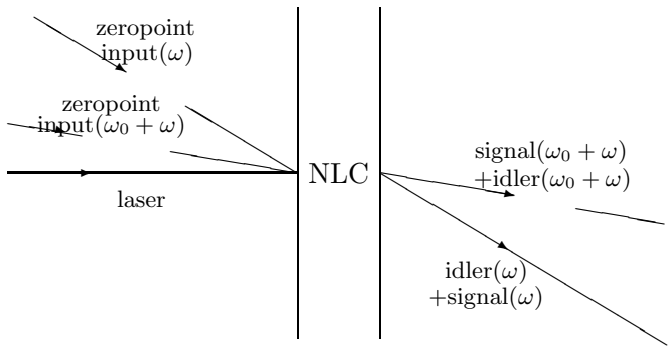
difference from the PDC situation, because
we find that the idler intensities are now less than the
input zeropoint intensities. The signal intensities in
both channels almost, but not quite, cancel this shortfall,
so that the PUC intensities, above threshold, are only about
3 per cent of the corresponding PDC quantities, which may explain
why nobody has yet observed them. Furthermore, the
intensity of the $\omega_0+\omega$ output is
actually less than the zeropoint input, and will
not therefore be detected at all.

But the intensity of the other output is above
zeropoint and my prediction, therefore, is that, as well as the main
PDC rainbow $\theta_d(\omega)$, {\it there is a satellite
rainbow}, whose intensity is about 3 percent of the
main rainbow, at $\theta_u(\omega)$. Although the precise
position of the satellite rainbow depends on the details
of the refractive index as a function of frequency, an
approximate calculation indicates\cite{puc1} that the component
of frequency $\omega_0/2$ will have a value for
$\theta_u$ about 2.5 times the corresponding
$\theta_d$.

\noindent
{\bf Acknowledgement}

\noindent
I have had a lot of help with the ideas behind this article,
and also in developing the argument, from Emilio Santos.

\end{document}